\documentclass[preprint2]{aastex}                                           
        
\shorttitle{Population clustering as a signal for deconfinement} 

\shortauthors{Poghosyan, Grigorian, Blaschke}

\begin{document} 

\title{Population clustering as a signal for deconfinement \\
in accreting compact stars}

\author{G.  Poghosyan\altaffilmark{1,3}, H.
Grigorian\altaffilmark{1}, D.  Blaschke\altaffilmark{2,3}}

\altaffiltext{1}{Department of Physics, Yerevan State University,
375025 Yerevan, Armenia}
\altaffiltext{2}{Fachbereich Physik, Universit\"{a}t Rostock, D-18051
Rostock, Germany}
\altaffiltext{3}{European Centre for Theoretical Studies in Nuclear 
Physics and Related Areas, Strada delle Tabarelle 286, 38050 
Villazzano (Trento), Italy}

\begin{abstract} 

We study the evolution of the rotation frequency for accreting compact
stars.  The discontinuous change of the moment of inertia of a rapidly
rotating star due to the possible quark core appearance entails a
characteristic change in the spin evolution.  Numerical solutions have
been performed using a model equation of state describing the
deconfinement phase transition.  Trajectories of spin evolution are
discussed in the angular velocity - baryon number plane ({\it phase
diagram}) for different accretion scenarios defined by the initial
values of mass and magnetic field of the star, as well as mass
accretion rate and magnetic field decay time.  We observe a
characteristic increase in the {\it waiting time} when a configuration
enters the quark core regime.  Overclustering of the population of Z
sources of LMXBs in the phase diagram is suggested as a direct
measurement of the critical line for the deconfinement phase transition
since it is related to the behaviour of the moment of inertia of the
compact star.

\end{abstract}



\section{Introduction} 

The deconfinement phase transition into a plasma of quasifree quarks
and gluons is expected to occure under conditions of sufficiently high
temperatures and/or densities as, e.g., in heavy-ion collisions, a few
microseconds after the big bang or in the cores of superdense stellar
objects.

For the latter, signals of a phase transition have been suggested in
the form of characteristic changes of observables \citep{fwbook}, such
as the pulse timing \citep{frido,cgpb}, brightness \citep{magnetar} and
surface temperature \citep{schaab,cool} of the isolated pulsars during
their evolution.

Recently, quasiperiodic brightness oscillations (QPOs) in low-mass
X-ray binaries (LMXBs) have been observed \citep{klis} which provide
new mass and radius constraints for compact objects \citep{MLP}.
According to the model of Stella and Vietri \citep{stella,stella2}
these constraints would allow only very compact objects made up of
strange quark matter \citep{strange}.  Due to the mass accretion flow
these systems are candidates for the most massive compact stars at the
limit to the formation of black holes.  Therefore, if the possible
deconfinement phase transition in compact stars exists at all, we
expect it to occur in accreting LMXBs \citep{cgpb,gw00}.

In this paper we study the spin evolution of compact stars with
possible quark matter cores for a disc accretion model and show that a
pronounced signal of a deconfinement transition in the star interior
can be obtained for accretors with long-lived magnetic fields and large
angular momentum transfer.  The so called Z sources of QPOs in LMXBs
are suggested as objects which should predominantly populate the region
of the suspected phase border between hadronic stars and quark core
stars (QCSs) \citep{cgpb}.  We suggest this population clustering as a
signal for deconfined quark matter in accreting compact stars.

\section{Evolution in the $\Omega-\uppercase{N}$ diagram}

Our investigation is based on the classification of rotating compact
star configurations in the plane of angular velocity $\Omega$ and
baryon number $N$, the so called {\it phase diagram} \citep{phdiag}.

We consider in our calculations a generic form of a relativistic
equation of state with a deconfinement transition \citep{cgpb}, which
describes the hadronic matter by a Walecka model and the quark matter
by a dynamical confinig model \citep{b+99,bt}.  The deconfinement phase
transition is obtained by imposing Gibbs' conditions for phase
equilibrium with the constraints that baryon number as well as electric
charge of the system are globally conserved \citep{glen}.  The onset of
deconfinement (mixed phase) is obtained at a density of about $1.5$
times the nuclear saturation density.

It has been shown that the critical line $N_{\rm crit}(\Omega)$, which
separates the region of QCSs from the hadronic ones, is correlated with
the local maxima of the moment of inertia with respect to changes of
the baryon number at given $\Omega$ due to the change of the internal
structure of the compact object at the deconfinement phase transition.
Therefore, we expect that the rotational behavior of these objects
changes in a characteristic way when this line is crossed.  The
consequence will be an increase of the population of stars at this
critical line which could be observed, provided that a sufficiently
large number of accretors will be detected with their masses and spin
frequencies \citep{phdiag}.

We consider the spin evolution of a compact star under mass accretion
from a low-mass companion star as a sequence of stationary states of
configurations (points) in the phase diagram spanned by $\Omega$ and
$N$.  The process is governed by the change in angular momentum of the
star

\begin{equation} 
\label{djdt} 
\frac{d}{dt} (I(N,\Omega)~ \Omega)= K_{\rm ext}~,
\end{equation}
where
\begin{equation}
K_{\rm ext}= \sqrt{G M \dot M^2 r_0}- N_{\rm out}
\label{kex} 
\end{equation} 
is the external torque due to both the specific angular
momentum transfered by the accreting plasma and the magnetic plus
viscous stress given by $N_{\rm out}=\kappa \mu^2 r_c^{-3}$, 
$\kappa=1/3$ \citep{lipunov}. For a star with radius
$R$ and magnetic field strength $B$, the magnetic moment is given by
$\mu=R^3~B$. The co-rotating radius
$r_c=\left(GM/\Omega^2\right)^{1/3}$ is very large ($r_c\gg r_0$)
for slow rotators.
The inner radius of the accretion disc is
\[ 
r_0 \approx \left\{
\begin{array}{cc} 
R~,&\mu < \mu_c \\ 
0.52~r_A~,&\mu \geq \mu_c 
\end{array} 
\right.
\] 
where $\mu_c$ is that value of the magnetic moment of the star for
which the disc would touch the star surface.  The characteristic
Alfv\'en radius for spherical accretion with the rate $\dot M=m \dot N$
is $r_A=\left(2\mu^{-4} G M \dot M^2\right)^{-1/7}$.  Since we are
interested in the case of fast rotation for which the spin-up torque
due to the accreting plasma in Eq.  (\ref{kex}) is partly compensated
by $N_{\rm out}$, eventually leading to a saturation of the spin-up, we
neglect the spin-up torque in $N_{\rm out}$ which can be important only
for slow rotators \citep{gl},

From Eqs.  (\ref{djdt}), (\ref{kex}) one can obtain the first order
differential equation for the evolution of angular velocity

\begin{equation} 
\label{odoto} 
\frac{d \Omega}{d t}= 
\frac{K_{\rm ext}(N,\Omega)- K_{\rm int}(N,\Omega)} 
{I(N,\Omega) + {\Omega}({\partial I(N,\Omega)}/{\partial \Omega})_{N}}~, 
\end{equation} 
where 
\begin{equation}
\label{kint}
K_{\rm int}(N,\Omega)=\Omega\dot N 
({\partial I(N,\Omega)}/{\partial N})_{\Omega}~. 
\end{equation}
 
Solutions of (\ref{odoto}) are trajectories in the $\Omega - N$ plane
describing the spin evolution of accreting compact stars, see Fig.
\ref{fig:spinning}.  Since $I(N,\Omega)$ exhibits characteristic
functional dependences \citep{phdiag} at the deconfinement phase
transition line $N_{\rm crit}(\Omega)$ we expect observable
consequences in the $\dot P - P$ plane when this line is crossed.

In our model calculations we assume that both the mass accretion and
the angular momentum transfer processes are slow enough to justify the
assumption of quasistationary rigid rotation without convection.  The
moment of inertia of the rotating star can be defined as $I(N,\Omega)=
J(N,\Omega)/\Omega~$, where $J(N,\Omega)$ is the angular momentum of
the star.  For a more detailed description of the method and analytic
results we refer to \citep{cgpb} and the works of
\citet{hartle,thorne}, as well as \citet{chubarian,sedrakian}.

The time dependence of the baryon number for the constant accreting
rate $\dot N$ is given by
\begin{equation} 
N(t)=N(t_0)+ (t-t_0)\dot N~.
\end{equation} 
For the magnetic field of the accretors we consider the exponential 
decay \citep{heuvel}
\begin{equation} 
B(t)=[B(0) - B_{\infty}]\exp(-t/\tau_B)+ B_{\infty}~.
\end{equation} 
We solve the equation for the spin-up evolution (\ref{odoto}) of the
accreting star for decay times $10^7\le \tau_B {\rm [yr]} \le 10^9$ and
initial magnetic fields in the range $0.2 \leq B(0){\rm [TG]}\leq 4.0
$.  The remnant magnetic field is chosen to be
$B_\infty=10^{-4}$TG\footnote[1]{1 TG= $10^{12}$ G} \citep{page}.

At high rotation frequency, both the angular momentum transfer from
accreting matter and the influence of magnetic fields can be small
enough to let the evolution of angular velocity be determined by the
dependence of the moment of inertia on the baryon number, i.e.  on the
total mass.  This case is similar to the one with negligible magnetic
field considered in \citep{shapiro,colpi,cgpb} where $\mu \leq \mu_c$
in Eq.  (\ref{odoto}), so that only the so called internal torque term
(\ref{kint}) remains.

In Fig.  \ref{fig:spinning} we show evolutionary tracks of accretors in
phase diagrams (left panels) and show the corresponding spin evolution
$\Omega(t)$ (right panels).  In the lower panels, the paths for
possible spin-up evolution are shown for accretor models initially
having a quark matter core ($N(0)= 1.55~N_\odot, ~\Omega(0)=1$ Hz).
The upper panels show evolution of a hybrid star without a quark matter
core in the initial state ($N(0)= 1.4~N_\odot,~\Omega(0)=1$ Hz),
containing quarks only in mixed phase.  We assume a value of $\dot N$
corresponding to observations made on LMXBs, which are divided into Z
sources with $\dot N \sim 10^{-8} N_\odot/$yr and A(toll) sources
accreting at rates $\dot N \sim 10^{-10} N_\odot/$yr
\citep{gw00,heuvel,klis}.

For the case of a small magnetic field decay time $\tau_B=10^7$ yr
(solid and dotted lines in Fig.\ref{fig:spinning}) the spin-up
evolution of the star cannot be stopped by the magnetic braking term so
that the maximal frequency consistent $\Omega_{max}(N)$ with stationary
rotation can be reached regardless whether the star did initially have
a pure quark matter core or not.

For long lived magnetic fields ($\tau_B=10^9$ yr, dashed and dot-dashed
lines in Fig.\ref{fig:spinning}) the spin-up evolution deviates from
the monotonous behaviour of the previous case and shows a tendency to
saturate.  At a high accretion rate (dot-dashed lines) the mass load
onto the star can be sufficient to transform it to a black hole before
the maximum frequency could be reached whereas at low accretion (dashed
lines) the star spins up to the Kepler frequency limit.

\section{Waiting time and population clustering}

The question arises whether there is any characteristic feature in the
spin evolution which distinguishes trajectories that traverse the
critical phase transition line from those remaining within the initial
phase.

For an accretion rate as high as $\dot N=10^{-8}~N_{\odot}/$ yr the
evolution of the spin frequency in Fig.  \ref{fig:spinning} shows a
plateau where the angular velocity remains within the narrow region
between $2.1 \le \Omega[{\rm kHz}]\le 2.3$ for the decay time
$\tau_B=10^9$ yr and between $0.4\le\Omega[{\rm kHz}]\le 0.5$ when
$\tau_B=10^7$ yr.  This plateau occurs for stars evolving into the QCS
region (upper panels) as well as for stars remaining within the QCS
region (lower panels).  This saturation of spin frequencies is mainly
related to the compensation of spin-up and spin-down torques at a level
determined by the strength of the magnetic field.  In order to perform
a more quantitative discussion of possible signals of the deconfinement
phase transition we present in Fig.  \ref{fig:PdotP} trajectories of
the spin-up evolution in the $\dot P - P$ plane for stars with
$N(0)=1.4~N_\odot$ and $\Omega(0)=1$ Hz in the initial state; the four
sets of accretion rates and magnetic field decay times coincide with
those in Fig.  \ref{fig:spinning}.

When we compare the results for the above hybrid star model (solid
lines) with those of a hadronic star model (quark matter part of the
hybrid model omitted; dotted lines) we observe that only in the case of
high accretion rate ($\dot N=10^{-8}N_\odot$/yr, e.g.  for Z sources)
and long-lived magnetic field ($\tau_B=10^9$yr) there is a significant
difference in the behaviour of the period derivatives.  The evolution
of a star with deconfinement phase transition shows a dip in the period
derivative in a narrow region of spin periods.  This feature
corresponds to a plateau in the spin evolution which can be quantified
by the {\it waiting time} $\tau=\left|P/\dot P\right|=\Omega/
\dot\Omega$.  In Fig.  \ref{fig:life} (lower and middle panels) we
present this {waiting time} in dependence on the rotation frequencies
$\nu=\Omega/(2\pi)$ for the relevant case labeled (9,-8) in Figs.
\ref{fig:spinning},\ref{fig:PdotP}.  The comparison of the trajectory
for a hybrid star surviving the phase trasition during the evolution
(solid line) with those of a star evolving within the hadronic and the
QCS domains (dotted line and dashed lines, respectfuly), demonstrates
that an enhancement of the waiting time in a narrow region of
frequencies is a characteristic indicator for a deconfinement
transition in the accreting compact star.

The position of this peak in the waiting time depends on the initial
value of magnetic field, see Fig.  \ref{fig:life}.  In the middle and
lower panels of that Figure, we show the waiting time distribution for
$B(0)= 0.75$ TG and $B(0)= 0.82$ TG, respectively.  Maxima of the
waiting time in a certain frequency region have the consequence that
the probability to observe objects there is increased ({\it population
clustering}).  In the upper panel of this Figure the spin frequencies
for observed Z sources in LMXBs with QPOs \citep{klis} are shown for
comparison.  In order to interprete the clustering of objects in the
frequency interval $225 \le \nu[{\rm Hz}] \le 375$ as a phenomenon
related to the increase in the waiting time, we have to chose initial
magnetic field values in the range $1.0\ge B(0)[{\rm TG}]\ge 0.6$ for
the scenario labeled (9,-8), see also the dashed lines in 
Fig.\ref{fig:PdotP}.

\section{Signal for Deconfinement in LMXBs}

The results of the previous section show that the waiting time for
accreting stars along their evolution trajectory is larger in a
hadronic configuration than in a QCS, after a time scale when the mass
load onto the star becomes significant.  This suggests that if a
hadronic star enters the QCS region, its spin evolution gets enhanced
thus depopulating the higher frequency branch of its trajectory in the
$\Omega - N$ plane.

In Fig.  \ref{fig:ONTcntr} we show contours of waiting time regions in
the phase diagram.  The initial baryon number is $N(0)=1.4 N_\odot$ and
the initial magnetic field is taken from the interval $0.2\leq
B(0)[{\rm TG}] \leq 4.0$ .

The region of longest waiting times is located in a narrow branch
around the phase transition border and does not depend on the evolution
scenario after the passage of the border, when the depopulation occurs
and the probability to find an accreting compact star is reduced.
Another smaller increase of the waiting time and thus a population
clustering could occur in a region where the accretor is already a QCS.
For an estimate of a population statistics we show the region of
evolutionary tracks when the values of initial magnetic field are
within $0.6\leq B(0)[{\rm TG}] \leq 1.0$ as suggested by the
observation of frequency clustering in the narrow interval $375 \geq
\nu [{\rm Hz}] \geq 225$, see Fig.  \ref{fig:life}.

As a strategy of search for QCSs we suggest to select from the LMXBs
exhibiting the QPO phenomenon those accreting close to the Eddington
limit \citep{heuvel} and to determine simultaneously the spin frequency
and the mass \citep{LM00} for sufficiently many of these objects.  The
emerging statistics of accreting compact stars should then exhibit the
population clustering shown in Fig.  \ref{fig:ONTcntr} when a
deconfinement transition is possible.  If a structureless distribution
of objects in the $\Omega - N$ plane will be observed, then no firm
conclusion about quark core formation in compact stars can be made.

For the model equation of state on which the results of our present
work are based, we expect a baryon number clustering rather than a
frequency clustering to be a signal of the deconfinement transition in
the compact stars of LMXBs.  The model independent result of our study
is that a population clustering in the phase diagram for accreting
compact stars shall measure the critical line $N_{\rm crit}(\Omega)$
which separates hadronic stars from QCSs where the shape of this curve
can discriminate between different models of the nuclear EoS at high
densities.

\acknowledgements

We thank M.  Colpi and N.K.  Glendenning for their stimulating interest
in our work; D.  Sedrakian, C.  Gocke and F.  Weber for their useful
remarks.  H.  G.  and G.  P.  acknowledge the hospitality of Rostock
University.  D.  B.  and G.  P.  are grateful for a Fellowship from the
ECT* Trento.  This work was supported in part by the {\sc Deutsche
Forschungsgemeinschaft} (DFG) under Grants No.  436 ARM 17/1/00 and 436
ARM 17/7/00, and by the {\sc Deutscher Akademischer Austauschdienst
(DAAD)}.

\clearpage

\plotone{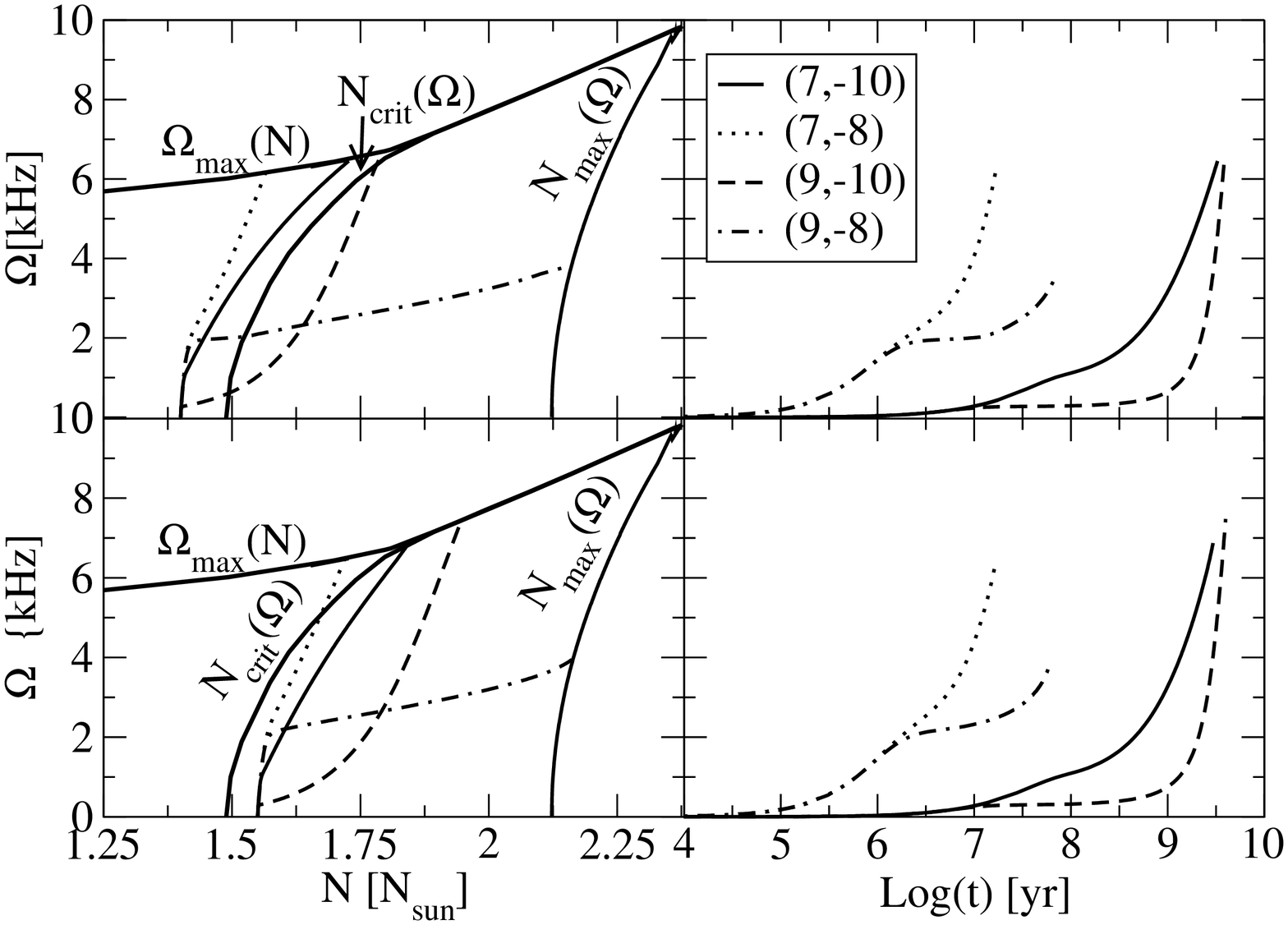}
\figcaption[spinningb.eps] {Spin evolution of an accreting compact star
for different decay times of the magnetic field and different accretion
rates.  Upper panels:  initial configuration with $N(0)=1.4~N_\odot$;
Lower panels:  $N(0)=1.55~N_\odot$; $\Omega(0)=1$ Hz in both cases.
The numbers in the legend box stand for ($\log (\tau_B [{\rm
yr}])$,~$\log (\dot N [N_\odot/{\rm yr}]$).  For instance (9,-8)
denotes $\tau_B=10^9$ yr and $\dot N=10^{-8}N_\odot/$ yr.
\label{fig:spinning}}

\plotone{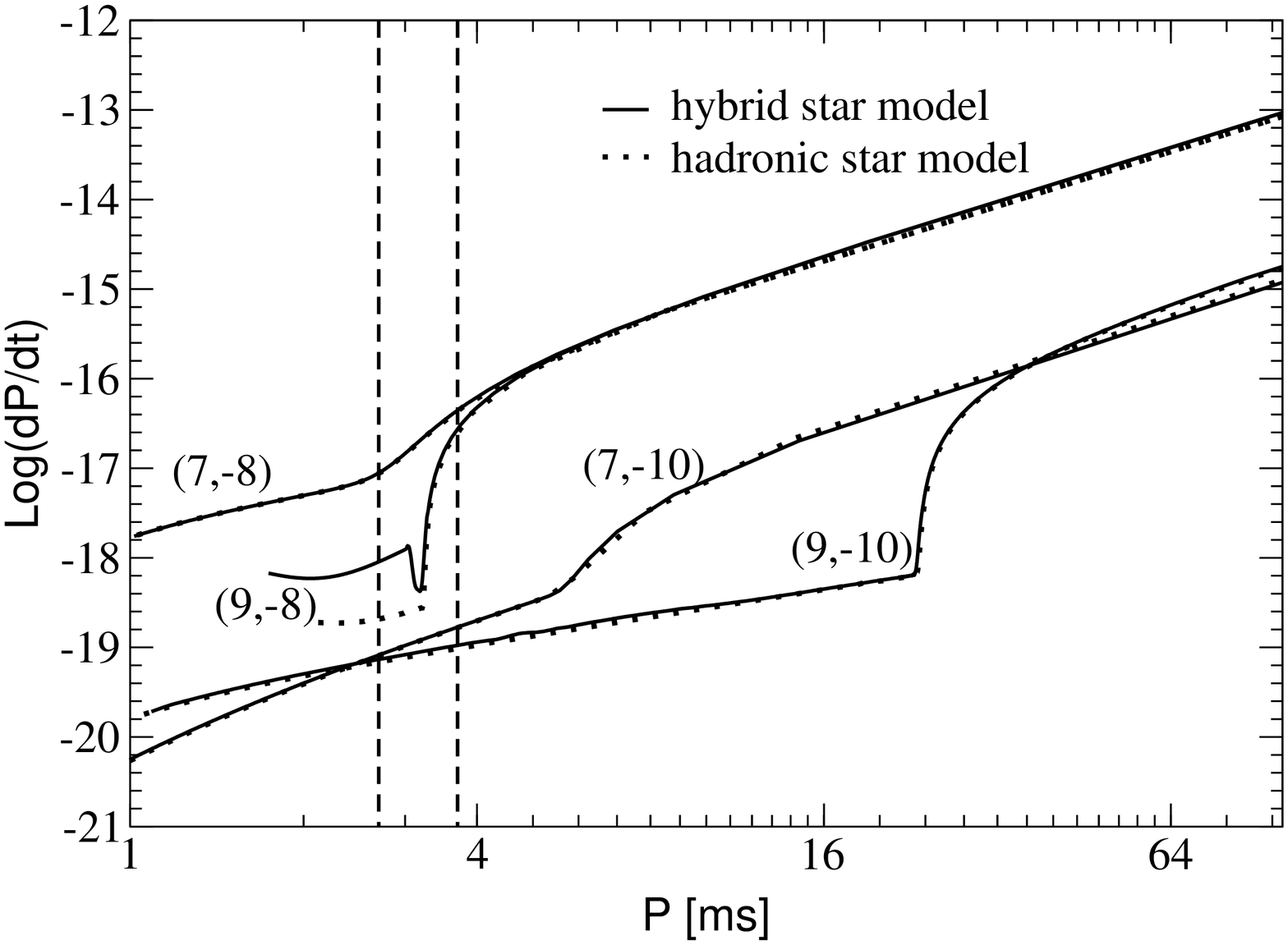}
\figcaption[PdotPb.eps] {Spin-up evolution of accreting compact stars
in the $\dot P-P$ diagram.  For labels and initial values see to Fig.
\protect\ref{fig:spinning}.  The region of the vertical dashed lines
corresponds to the clustering of periods observed for LMXBs with QPOs.
A dip (waiting point) occurs at the deconfinement transition for
parameters which correspond to Z sources ($\dot N=10^{-8} N_\odot$/yr)
with slow magnetic field decay ($\tau_B=10^9$yr).  \label{fig:PdotP}}

\plotone{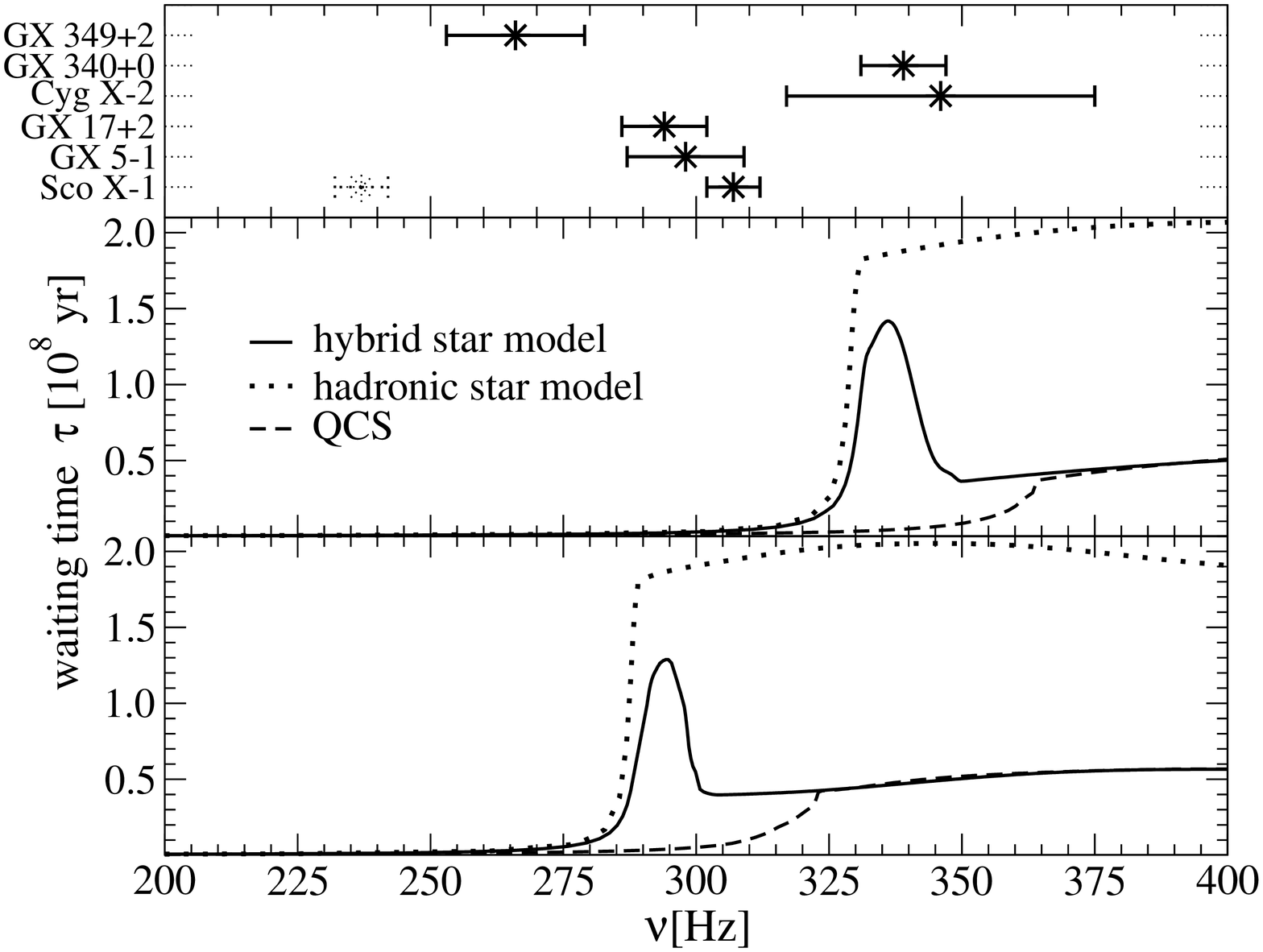}
\figcaption[liferb.eps] {{\it Upper panel:}  frequency interval for
observed Z source LMXBs \protect\citep{klis}; {\it middle panel:}
waiting times $\tau=P/\dot P$ for scenario (9,-8) and initial magnetic
field $B(0)= 0.75$ TG; {\it lower panel:}  same as middle panel for
$B(0)= 0.82$ TG.  Spin evolution of a hybrid stars (solid lines) shows
a peak in the waiting time characteristic for the deconfinement
transition.  Hadronic stars (dotted lines) and QCSs (dashed lines) have
no such structure.  \label{fig:life}}

\plotone{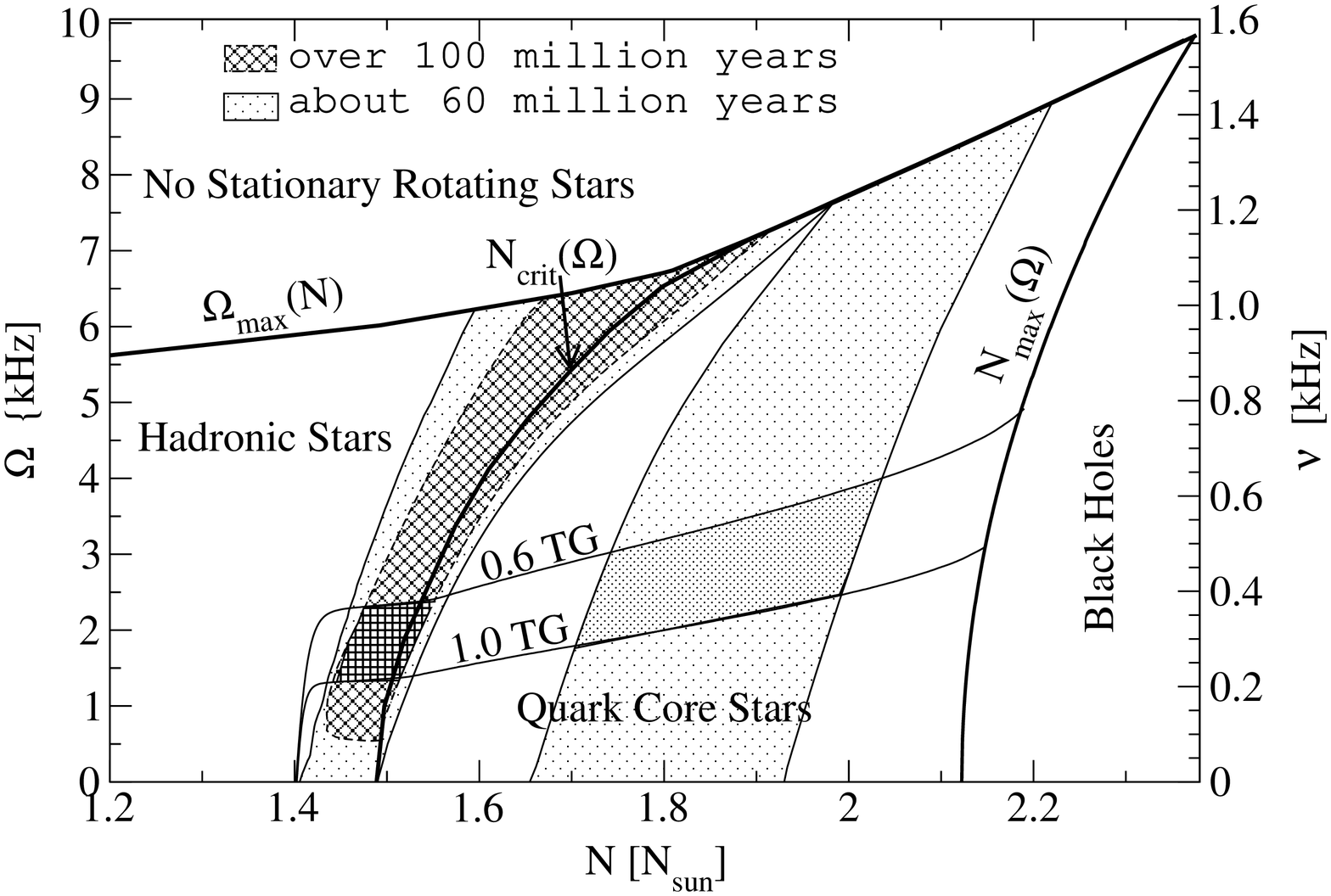}
\figcaption[ontcontrb.eps] {Regions of {waiting times} in the phase
diagram for compact hybrid stars for the (9,-8) scenario.  For an
estimate of a population statistics we show the region of evolutionary
tracks when the interval of initial magnetic field values is restricted
to $0.6\leq B(0)[{\rm TG}] \leq 1.0$.  Note that the probability of
finding a compact star in the phase diagram is enhanced in the vicinity
of the critical line for the deconfinement phase transition $N_{\rm
crit}(\Omega)$ by at least a factor of two relative to all other
regions in the phase diagram.  \label{fig:ONTcntr}}

\end{document}